\documentclass[11pt]{article}

\usepackage{fullpage, hyperref,bbm,amssymb,amsfonts,amsthm}
\usepackage[noend]{algorithmic}
\usepackage{algorithm}
\usepackage{setspace}
\usepackage{color}
\usepackage{latexsym}
\usepackage{verbatim} 
\usepackage{graphics}
\usepackage{graphicx}

\renewcommand{\>}{\rangle}

\newtheorem{theorem}{Theorem}

\begin{document}
\title{Quantum Phase Estimation with an Arbitrary Number of Qubits} 
\author{ 
Chen-Fu Chiang\thanks{ D\'{e}partement de Physique, Universit\'{e} de Sherbrooke Sherbrooke,
Qu\'{e}bec, Canada J1K 2R1S Email:
\texttt{Chen-Fu.Chiang@USherbrooke.ca}}
}		
\date{}		
\maketitle

\begin{abstract}
Due to the great difficulty in scalability, quantum 
computers are limited in the number of qubits during the early stages of the
quantum computing regime. In addition to the required qubits for
storing the corresponding eigenvector, suppose we have additional $k$ qubits
available. Given such a constraint $k$, we propose an approach for the phase estimation for an eigenphase of exactly $n$-bit precision. 
This approach adopts the standard recursive circuit for quantum Fourier transform (QFT) in \cite{CW:00} and
adopts classical bits to implement such a task. Our algorithm has
the complexity of $O(n \log k)$, instead of $O(n^2)$ in the conventional
QFT, in terms of the total invocation of rotation gates.
We also design a scheme to implement the factorization algorithm by using $k$
available qubits via either the continued fractions approach or the simultaneous
diophantine approximation. 
\end{abstract}

\section{Introduction}
Quantum phase estimation (QPE) is a key quantum operation in many quantum
algorithms \cite{Hallgren:02, Shor:94, Shor:97, Szegedy:04, WCNA:09}. Phase estimation 
is extensively used to solve a variety of problems, such as
hidden subgroup, graph isomorphism, quantum walk, quantum sampling, adiabatic
computing, order-finding and large number factorization. QPE
comprises two components: {\em phase kick back} and {\em inverse quantum
Fourier transform}. The implementation of quantum Fourier transform has been described in numerous research
articles \cite{CW:00, KSV:02, NC:00, HC:11, Cheung:04}. The physical
implementation (algorithms based on quantum Fourier transform (QFT)) is highly constrained by the requirement of 
(1) high-precision controlled rotation gates (phase shift operators), which remain difficult to realize, and (2) sufficient 
number of qubits to approximate the eigenphase to a required precision.
\\

At the early stage of a quantum computing implementation, we can
imagine that scalability could be an issue. The quantum resources could be
limited, in terms of available quantum qubits and quantum gates. From that
perspective, efficient implementations of quantum algorithms are essential 
when available quantum resources are scarce. For instance, Parker and Plenio \cite{PP:00}
show that a single pure qubit together with a collection of $\log_2 N$ qubits in
an arbitrary mixed (or pure) state is sufficient to implement Shor's
factorization algorithm efficiently to factorize a large number $N$. Such
implementation addresses the issue of limited qubits but introduces the concern
for the decoherence. \\

In this paper, we are interested in the following two aspects. (1) Given certain
available qubits, assuming $k + \log_2 N $ qubits in total, we want to have an
efficient way to implement quantum phase estimation and use as few
controlled rotation gates (c-r.g.) as possible. (2) Apply this technique to
Shor's factorization algorithm along with simultaneous diophantine approximation \cite{Seifert:01} to investigate the feasible implementation structure when the available qubits are limited. We
assume only one copy of the eigenvector $|u\>$ (requiring $\log_2 N$ qubits) and
additional $k$ qubits are available. One copy of the eigenvector
implies a restriction on the use of controlled-U gates: all controlled-U
gates should be applied on the workspace register ($k$ qubits). \\

One copy of an eigenvector is a reasonable assumption because multiple copies
of $|u\>$ would imply the requirement for extra multiple of $\log_2 N$ qubits for storage.
Hence, it is practical as we are considering the case that the available
qubits are scarce. Thus, the entire process is a single circuit ($\lceil n/k
\rceil$ stages) that {\em can not be divided into parallel processes}. Under
such an assumption, for approaches that require repetitions, such as Kitaev's
\cite{KSV:02} and others \cite{HC:11}, parallelization can not be done and the circuit depth is the
same as the size of the circuit. On the other hand, if we have enough
qubits for storing multiple copies of eigenvector $|u\>$, we should
choose Kitaev's approach because the processes can thus be run in parallel. Throughout the rest of the article,
we will refer to the $k$ available qubits as the qubits used in the workspace register.\\

Generally speaking, quantum circuits for QFT implemented in different approaches
\cite{CW:00, KSV:02, NC:00, HC:11, Cheung:04} would require the same number of controlled-U gates but different numbers of
rotation gates. We are interested in using the recursive approach, along with
some classical resources, to implement the inverse quantum Fourier transform. We
bound the number of required rotation gates from above. 
\\

We give an overview of the conventional quantum phase estimation technique in
section \ref{sect:overview}. We detail our algorithms and the analysis in
section \ref{sect:theAlgorithm}, including a brief analysis of
Kitaev's original approach \cite{KSV:02}. An application of our approach along
with simultaneous diophantine approximation to the factorization problem is given in section
\ref{sect:application}. Finally we state our conclusion in section
\ref{sect:theConclusion}.
\section{Approach based on QFT}\label{sect:overview}

\begin{figure}[ht!]
	  \begin{center}
		\includegraphics[height=1.3in,width=3.7in]{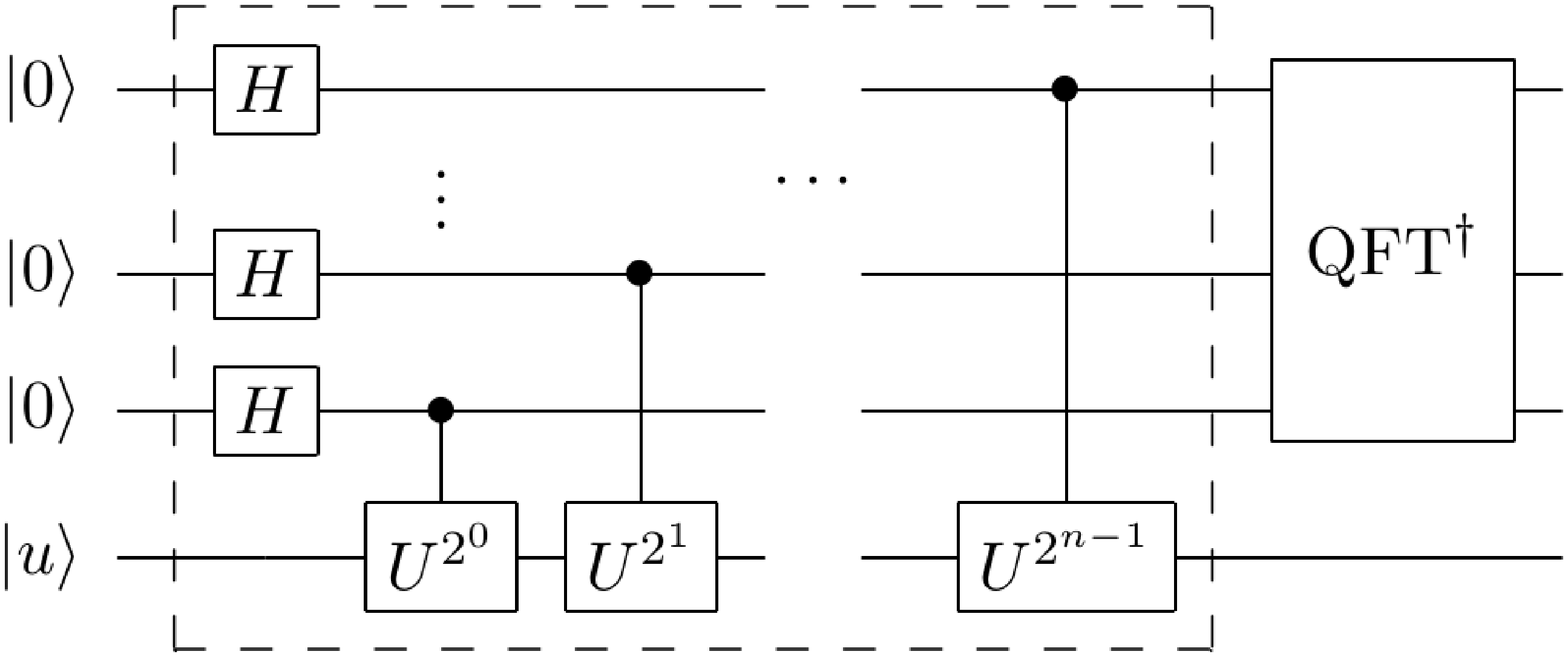} 
	  \end{center}
	 \caption {Standard QPE with $n$ qubits as ancilla.
	 The dash-line box is the phase kickback.}
	 \label{QPEfig}
	\end{figure}

One of the standard methods to approximate the phase of a unitary matrix is QPE based on QFT. The structure of this method
is depicted in Figure~\ref{QPEfig}. 

The QPE algorithm requires two registers and contains two stages. Suppose the
eigenphase of unitary $U$ is $\varphi = 0.\varphi_1 \varphi_2 \ldots \varphi_n$
in the binary representation such that
\begin{equation}\label{eqn:phaseestimate}
U|u\>=e^{2\pi i \varphi}|u\>.
\end{equation}
Then the first register is  prepared as a
composition of $n$ qubits initialized in the state $|0\>$. The second register is initially prepared in 
the state $|u\>$. The first stage prepares a uniform superposition over all
possible states and then applies controlled-$U^{2^l}$  operations. Consequently, the state becomes
\begin{equation}\label{stateStage1}
\frac{1}{2^{n/2}}\sum_{l=0}^{2^n-1}e^{2 \pi i \varphi l}|l\>. 
\end{equation}

The second stage in the QPE algorithm is the QFT$^\dag$ operation. At each
step (starting from the least significant bit) by using the information from
previous steps, the inverse Fourier transform transforms the state
\begin{equation}
\frac{1}{\sqrt{2}}(|0\>+e^{2\pi i 2^l\varphi}|1\>)
\end{equation}
to get closer to one of the states 
\begin{equation}
\frac{1}{\sqrt{2}}(|0\>+e^{2\pi i
0.0}|1\>) = \frac{1}{\sqrt{2}}(|0\> + |1\>) \quad \mbox{or} \quad
\frac{1}{\sqrt{2}}(|0\> + e^{2\pi i 0.1}|1\>) =
\frac{1}{\sqrt{2}}(|0\>-|1\>).
\end{equation}

\begin{figure}
 \begin{center}
    \includegraphics[height=1.3in,width=3.7in]{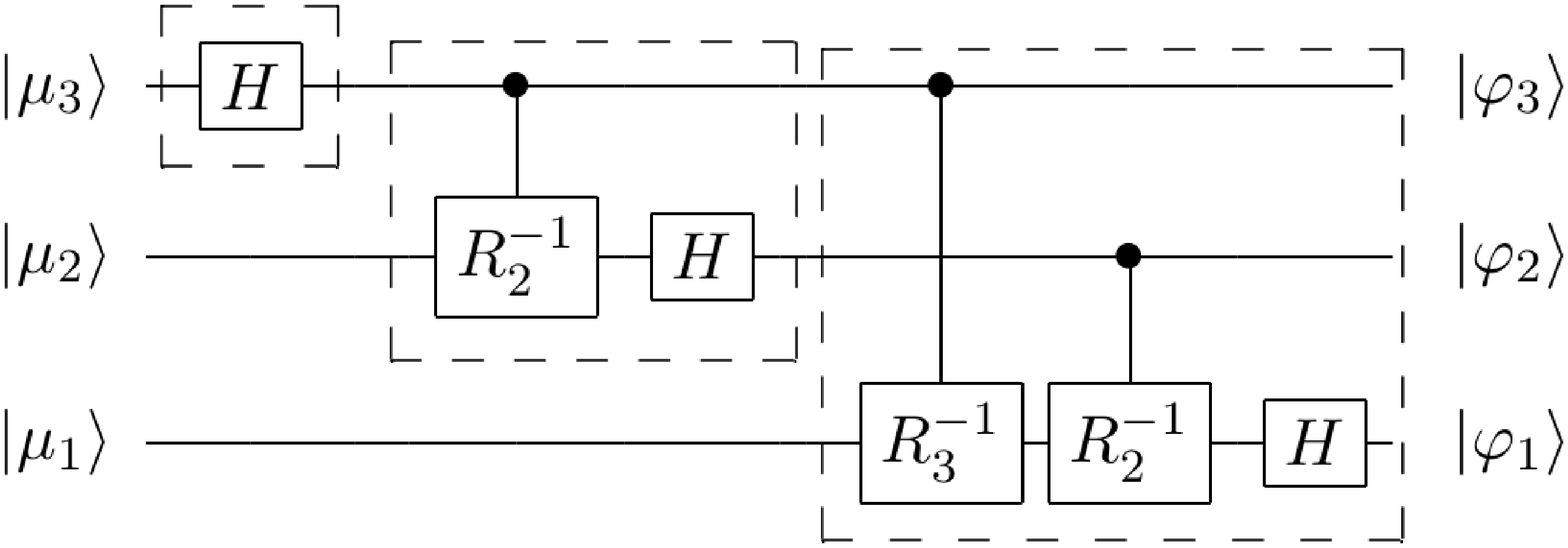} 
 \end{center}

\caption{3-qubit inverse QFT where $1 \leq j \leq 3$,
$|\mu_j\>=\frac{1}{\sqrt{2}}(\left|0\right>+e^{2\pi i(0.\varphi_j\ldots
\varphi_3)}\left|1\right>$). }
  \label{fig:InverseQFT_3qubit}
\end{figure}

Suppose $\varphi$ is precise to the $3$rd bit, that is $\varphi = 0.\varphi_1 \varphi_2 \varphi_3$. As shown in Figure \ref{fig:InverseQFT_3qubit}, 
each step (dashed-line box) uses the result of previous steps, where phase shift operators are defined as
\begin{equation}\label{eq:phaseshift}
R_l \equiv
\left[ {\begin{array}{cc}
 1 & 0  \\
 0 & e^{2\pi i/2^l}  \\
 \end{array} } \right] 
\end{equation}
 for $2\leq l \leq 3$. 
 By concatenating $\varphi_1$, $\varphi_2$ and $\varphi_3$, 
 we obtain $\varphi$. Therefore, when $\varphi$ is precise to the $n_{th}$ bit, the total number of rotation gate 
 invocations is $O(n^2)$.

\section{Our Algorithm}\label{sect:theAlgorithm}
Before proceeding to our algorithm, we provided the description of the recursive
circuit for quantum Fourier transform \cite{CW:00} technique. In \cite{CW:00},
in addition to the recursive circuit, the authors also adopted the technique by
Sch\"{o}nhage and Strassen \cite{SS:71} for integer multiplication. However,
the integer multiplication is performed via classical
computation. We use the classical bits and operators to compute the parameter (the desired phase shift) of a
quantum rotation gate. The algorithm structure is explained in subsection \ref{sect:algStruct}. 
\\

\subsection {Standard recursive circuit description for $F_{2^n}^\dagger$}
Let $F_{2^n}^\dagger$ denote the inverse Fourier transform modulo $2^n$ that
acts on $n$ qubits. The standard quantum circuit for $F_{2n}^\dagger$ can be described recursively as follows. Let us denote this circuit as $RF_{2^n}^\dagger$. 

\begin{enumerate}
  \item Suppose the state of the work register after phase kickback is 
  \begin{equation}
		|\psi\> = \frac{1}{\sqrt{2}}(|0\> + e^{2\pi i0.\varphi_1 \ldots \varphi_n}|1\>)\otimes \ldots \otimes \frac{1}{\sqrt{2}}(|0\> + e^{2\pi i0.\varphi_n}|1\>) = |\mu_1\>\otimes \ldots \otimes |\mu_n\>
	\end{equation} 
  \item Apply $F_{2^m}^\dagger$ to the last m qubits ($|\mu_{n-m+1}\> \otimes \ldots \otimes |\mu_n\>$).
  \item Read out and store the values of the $m$ qubits in classical bits
  ($c_{1}\ldots c_{m}$).
  \item Compute rotation angle: $f(c_{1}\ldots c_m) = \sum_{i=1}^m
  (\frac{1}{2})^{i}\cdot c_i$.
  \item For each $j \in \{ 1, 2, \ldots, n-m\}$, 
  apply the rotation gate $R_{\frac{f(c_1\ldots c_m)}{2^{n-m-j+1}}}^\dagger$ to the $j^{th}$
  qubit. Here the rotation gate  $R_{\frac{f(c_1\ldots c_m)}{2^{n-m-j+1}}}^\dagger$ is defined as 
  \[
  	R_q^\dagger = \left( \begin{array}{cc}
	1 & 0 \\
	0 & e^{-2\pi i \frac{f(c_1\ldots c_m)}{2^{n-m-j+1}}}  \end{array} \right)
  \]
  \item Apply $F_{2^{n-m}}^\dagger$ to the first $n-m$ qubits. 
\end{enumerate}
	For simplicity, let us assume that $n$ is some power of 2. Then step 5 is the 
	step that resets the disturbing eigenphase
	bits for the first $n-m$ qubits because all the disturbing eigenphase bits from
	the last $m$ bits will be cleared. The number of required rotation operations for
	such a step is $n/2$ (suppose we choose $m=n/2$) as we have to reset for each
	qubit in the last $n-m$ qubits. \\
	
	\noindent
	It is clear to that the total number of
	required rotation gates is
	\begin{equation}\label{eqn:recursive}
		T_n = T_{n/2} + T_{n/2} + n/2 
	\end{equation}
	where $T_1 = 1$. Hence, the complexity is $O(n \log n)$ \footnote{In this work, $\log$ is always of
base $2$, unless otherwise specified.} for such a recursive
	circuit.

\subsection{The algorithm structure} \label{sect:algStruct}

Given $k$ ancillary bits initialized in $|0\>$ and eigenvector $|u\>$ of unitary
$U$ where $U|u\> = e^{2 \pi i \varphi}|u\>$ as input, we want to estimate the
eigenphase of $U$ precise to the $n_{th}$ bit. The algorithm comprises $\lceil
n/k \rceil $ stages that run in sequence. At each stage, we perform phase
kickback, controlled-rotation operation and recursive inverse Fourier
transform to obtain $k$ eigenphase bits. Once the last stage finishes, we can
concatenate the obtained eigenphase bits, resulting in an estimated eigenphase of $\varphi$. For the details, please refer to Algorithm
\ref{alg:KQubits} listed below.

\begin{algorithm}[H]\caption{Phase Estimation with
Variable Number of qubits}\label{alg:KQubits}
\textbf{Input:} 
			$k$ ancillary bits initialized in $|0\>$ and eigenvector $|u\>$ of unitary
			$U$ where $U|u\> = e^{2 \pi i \varphi}|u\>$. \\
    		
    		\textbf{Step I:}\\
    		At stage $j$, where $j \in \{1, \ldots, \lceil n/k \rceil \}$, run phase kick back on $k$ qubits by using the 
    		controlled $U^{2^l}$ operations. 
    		Note that $l \in \{{n-jk},{n-jk+1},\ldots, {n-(j-1)k-1}\}$.  \\ 
    		
    		\textbf{Step II:} \\
    		For $t \in \{1, \ldots, k\}$, apply the rotation gate $R_{\frac{
    		F[j-1]}{2^{k-t+1}}}^\dagger$ to the $t_{th}$ qubit.\\
    		Apply the generalized recursive circuit $F_{2^k}^\dagger$. \\ 
    		Read out the result to $k$ classical bits $(c_1\ldots c_k)$ (the actual label is $c_{n-jk+1}\ldots c_{n-(j-1)k}$).\\
    		Compute the value $F[j] = f(c_1\ldots c_k) + \frac{F[j-1]}{2^k}$ where
    		$f(c_{1}\ldots c_k) = \sum_{i=1}^k (\frac{1}{2})^{i}\cdot c_i$. \\
    		Reset $k$ qubits to $|0\>$ \\ 
    		
            \textbf{Step III:} \\
            Repeat Step I and Step II $ \lceil n/k \rceil $ times (i.e. $\lceil n/k \rceil$ stages)\\ 
            
            \textbf{Output:} \\
            Concatenate the $n$ classical bits $c_1, \ldots , c_n$,
            resulting in an estimated eigenphase $\varphi = 0.c_1c_2c_3\ldots$.
	\end{algorithm}  
	\noindent
	Let us write the eigenphase $\varphi$ in the binary presentation as $0.\varphi_1\ldots \varphi_n$. Let $|\psi\> = |0\>^{\otimes k}|u\>$ be 
	the initial state at stage $j$ before the phase kickback. After Step I, we
	obtain the state
	\begin{equation}
		|\Phi\>_1 = \frac{1}{\sqrt{2}}(|0\> + e^{2\pi i0.\varphi_{n-jk+1}\ldots
		\varphi_{n-1} \varphi_n}|1\>)\otimes \ldots \otimes \frac{1}{\sqrt{2}}(|0\> + e^{2\pi i0.\varphi_{n-jk+k}\ldots \varphi_{n-1}\varphi_n}|1\>).
	\end{equation} 
	
	It is clear to see that for the $t_{th}$ qubit that the eigenphase discovered from previous stages is 
	shifted to the right by $k-t+1$ bits in the binary presentation. 
	At the beginning of Step II, by applying the rotation gate
	$R_{\frac{1}{2^{k-t+1}}\cdot F[j-1]}^\dagger$ \footnote{$F[0] = 0$.}, 
	we reset the discovered eigenphase in those $k$ qubits. Hence, we obtain the state
	\begin{equation}
		|\Phi\>_{2-1} = \frac{1}{\sqrt{2}}(|0\> + e^{2\pi i0.\varphi_{n-jk+1}\ldots \varphi_{n-jk+k}}|1\>)\otimes \ldots \otimes \frac{1}{\sqrt{2}}(|0\> + e^{2\pi i0.\varphi_{n-jk+k}}|1\>).
	\end{equation}
	Now we have reduced the scenario to the case where the disturbing eigenphase
	bits from previous stages are reset to $0$. Hence we can use the general
	recursive circuit for the inverse quantum Fourier transform to obtain the eigenphase bits ($\varphi_{n-jk+1}, \varphi_{n-jk+2}, \ldots , \varphi_{n-jk+k}$). \\
	
\begin{figure}[htb]
		 \begin{center}
		   \includegraphics[height=1.8in,width=5.6in]{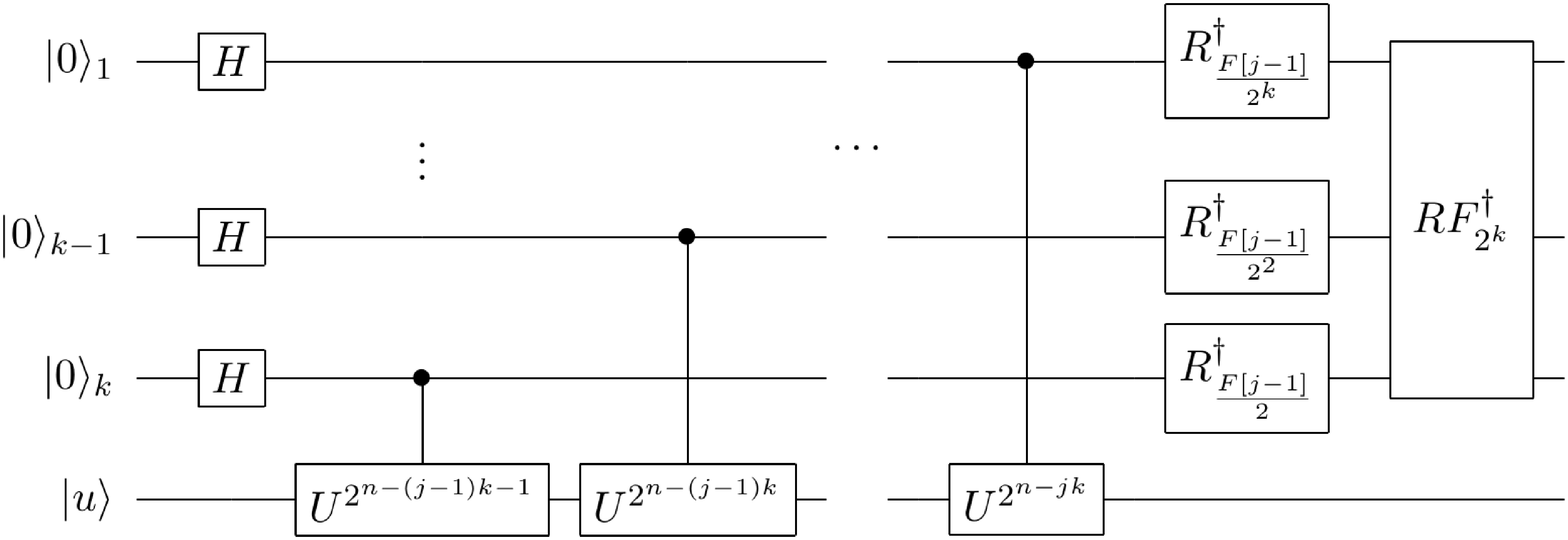} 
	  \end{center}
	 \caption {At stage $j$: with $k$ qubits as ancilla, $k$ rotation operations
	 and one $RF_{2^k}^\dagger$ operation.}
	 \label{fig:stage}
	\end{figure}

	Once we obtain the $k$ eigenphase bits, we can read out and store them in
	classical bits to compute $F[j]$. We refer interested readers to \cite{GN:96} for the details in this semiclassical approach. 
	The value, $F[j]$ will be used again in the
	next stage for resetting the previous $j \times k$ eigenphase bits. Figure
	\ref{fig:stage} depicts the process of a single iteration.

\subsection{The Analysis}\label{sect:theAnalysis}
	\noindent
	The cost of our algorithm has two parts: classical and quantum. 
	For the classical part, we need $n$ classical bits, $k+2$ doubles
	\newcounter{fnnumber}
	\footnote{Assume a classical double data
	structure is of size 64 bits.}%
	\setcounter{fnnumber}{\thefootnote}%
	  and $k+2$ classical operators. $n$ classical bits are used to store {\em all
	 of the} observed eigenphase bits. At any given stage (say $j$), two primitive
	 doubles\footnotemark[\thefnnumber], $X$ and $Y$ are required such that we
	 have
	\[
		X = F[j] , \quad Y = F[j-1].
	\]
	
	\noindent
	To generate $k$ different rotation angle operators (see the first substep
	of Step II in Algorithm \ref{alg:KQubits}), we need $k$ doubles ($Reg[k]$, an
	array of $k$ doubles) and $k$ operators \footnote{Because we can generate those
	parameters in parallel.} to generate the parameter,
	\[
		{\frac{ F[j-1]}{2^{k-t+1}}}, 
	\]
	of a quantum rotation gate for the $t_{th}$ qubit at the $j_{th}$ iteration
	where $t \in \{1,2, \ldots , k\}$.
	
	Once the $k$ eigenphase bits are stored in classical bits in
	the $j_{th}$ iteration, a classical operator computes $F[j]$ such that
	\[ 
	F[j] = X= f(c_1\ldots c_k) + \frac{Y}{2^k}.
	\] Then another operator sets $Y = X$. By doing so, double $X$ and $Y$ can
	be reused in the next iteration. Therefore, classically $n$ classical bits,
	$k+2$ doubles and $k+2$ classical operators are needed. The same device
	(classical requirement) can be used inside the recursive circuit since our
	approach is sequential, not parallelled. The classical requirements are summed in 
	Table \ref{table:bitNum}\footnotemark[\thefnnumber]. \\
	
	\begin{table}[tb]
	\caption{Required classical bits}
		\centering
			\begin{tabular}{|l|l|}
				\hline
					Register Type & Required number of bits \\
					\hline
					$X$ (classical register)                 & 64 \\
					$Y$ (classical register)                 & 64 \\
					Reg$[k]$ (classical register)			 & 64$k$ \\
					Classical bits for eigenphase            & $n$ \\
				    \hline
			\end{tabular}
	\label{table:bitNum}
	\end{table} 
	
	For the quantum part, the number of total rotation gate invocations in our
	approach would be
		\begin{equation}\label{eqn:invkcost}
		k \log k + (\lceil n/k
		\rceil -1)((k + k \log k)) \approx O(n \log k).
		\end{equation}
	The reasoning is as follows. At stage $j = 1$, the rotation operations only
	occur inside the recursive inverse Fourier transform $RF_{2^k}$ as $F[0] =0$. For stage $j = 2, \ldots ,
	\lceil \frac{n}{k} \rceil$, it is required to have rotation gates
	$R^\dagger_{\frac{F[j-1]}{2^{k-t+1}}}$, where $1 \leq t \leq k$, to reset the
	dangling eigenphase bits before the recursive inverse Fourier transform
	$RF_{2^k}$. Based on the cost fuction for $RF_{2^k}$ derived in Eqn.
	\ref{eqn:recursive}, we obtain the cost for our approach as shown in
	Eqn.~\ref{eqn:invkcost}.\\
	
	For comparison with other known existing
	approaches, in the following section we will briefly describe the analysis and the result rendered in \cite{HC:11} regarding Kitaev's 
	original approach \cite{KSV:02}.
		
	\subsubsection{Kitaev's Original Approach}\label{sect:theComparisonk}
\begin{figure}[htbp]
	 \begin{center}
		\includegraphics[height=0.9in,width=3in]{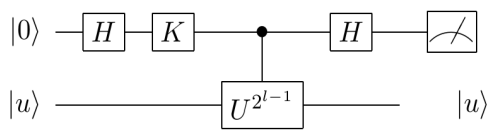} 
	  \end{center}
%
	\caption{Hadamard test with extra phase shift operator.} \label{QPE_with_K_Operator}
	\end{figure}

	\noindent
	In this approach, a series of Hadamard tests are performed for each eigenphase
	bit in order to recover the phase correctly. Suppose the precision up to the
	$n$th bit is required, then in each test the phase \footnote{see section \ref{sect:overview} for the description of the
	eigenphase $\varphi$ and the unitary $U$.} $\phi_l =
	2^{l-1}\varphi$ ($1\leq l\leq n$) must be computed up to precision
	$1/16$. We perform the Hadamard test on the $l_{th}$ eigenphase bit, starting
	from $l = n$ down to $1$, as depicted in
	Figure~\ref{QPE_with_K_Operator}.

	\noindent 	
	When $K = I_2$, the
	probabilities of post-measurement of the Hadamard test are 
	\begin{equation}\label{Eq:P1}
	\Pr(0|k)= \frac{1 + \cos(2\pi \varphi_k)}{2}, \quad 
	\Pr(1|k) = \frac{1 - \cos(2\pi \varphi_k)}{2}. 
	\end{equation}
	However, a cosine cannot distinguish $\phi_l$ and
	$-\phi_l$. We need to choose $ K = \left( {\begin{array}{cc}
	1 & 0 \\ 0 & i \\ \end{array} } \right)$ to be able to distinguish. The
	probabilities of the post-measurement states based on the modified Hadamard test become
	\begin{equation}\label{Eq:P2}
	\Pr(0|l)=\frac{1 - \sin(2\pi \phi_l)}{2}, \quad 
	\Pr(1|l) = \frac{1 + \sin(2\pi \phi_l)}{2}. 
	\end{equation}
	We then can recover $\phi_l$ from the estimates of the probabilities.
	To obtain the required precision of $1/16$ for $\phi_l$, we can run an iteration of Hadamard tests to estimate
	$\Pr(1|l)$ to some precision.
	
	\begin{theorem}\cite{HC:11}
	Assume $U$ is a unitary matrix with eigenvalue $e^{2\pi i \varphi}$ and
	corresponding eigenvector $|u\>$. Suppose $\varphi = 0.\varphi_1\ldots
	\varphi_n$ and let $\phi_l = 2^{l-1}\varphi$ ($1\leq l\leq n$). To obtain the
	required precision of $1/16$ for $\phi_l$ such that the recovered
	$\tilde{\varphi}$ is precise to the $n$th bit with constant success probabilty
	greater than $\frac{1}{2}$, for each $\phi_l$ we need to run at least $55 \ln
	n$ trials of Hadamard tests when using Kitaev's approach.
	\end{theorem}
	We refer the interested readers to \cite{HC:11} for the details. Since
	we have $n$ stages for $\phi_l$, the required invocation of a rotation gate
	(Hadamard in this case) in Kitaev's approach is $O(n \ln
	n)$. Suppose the controlled-rotation gates are precise, we list the
	comparison between Kitaev's approach, the conventional QFT based approach and our approach in Table  \ref{table:qbitNum}. \begin{table}[htb]
		\caption{The number of quantum rotation gates invocations}
			\centering
				\begin{tabular}{|l|l|l|l|}
					\hline
						Approach Type & Conventional & Kitaev's & Ours \\
						\hline
						Complexity & $O(n^2)$ & $O(n \ln n)$ & $O(n \log k)$ \\
						\hline
				\end{tabular}
	\label{table:qbitNum}
	\end{table}
\subsection{An Application}\label{sect:application}
In this section, we will focus on how to use $k$ available qubits to
implement the quantum factorization algorithm. Shor's factorization
algorithm provides a polynomial approach to factorize a large number $N$.
Suppose $N$ is an $L$ bit composite number of interest. There is no
known classical algorithm for factoring in only polynomial time, i.e., that can 
factor in time $O(L^c)$ for some constant $c$. The most difficult
integers to factor in practice using existing algorithms are those that are products of 
two large primes of {\em similar size}, and for this reason these are the
integers used in cryptographic applications. The largest such semiprime yet factored was RSA-768, a 768-bit number with 232
decimal digits \cite{Kleinjung:10}. \\

Quantumly, it is shown such a task can be done by using $O(L^3)$ operations. The
algorithm is two-fold. It first runs phase estimation to obtain the eigenphase 
$\varphi = 0.\varphi_1\varphi_2\ldots \approx s/r$ where $r$ is the order of
an arbitrary element $x$ (that is $x^r = 1 ($mod $N)$). The second part of the
algorithm involves the continued fractions algorithm to approximate $s/r$, based
on the eigenphase we obtain in phase estimation, in order to recover the order $r$. If $r$ is even, then
we know that $(x^{r/2} + 1)(x^{r/2}-1) = 0 ($mod $N)$ and we successfully factorize 
$N$ into a product of two large numbers of similar size. \\

However, using the continued fraction algorithm leads inevitably to
a squaring of the number to be factored. This follows from the following
theorem. 

\begin{theorem}\cite{NC:00}
Suppose $s/r$ is a rational number such that 
\[
	\big|\frac{s}{r} - \varphi \big| \leq \frac{1}{2r^2}.
\]
Then $s/r$ is a convergent of the continued fraction for $\varphi$, and thus can
be computed in $O(L^3)$ operations using the continued fractions algorithm.
\end{theorem}
This in turn doubles the length, approximately to $2L+1$ qubits, of the quantum registers in order to 
achieve required precision $1/2r^2$ since $ 1 \leq r \leq N \leq 2^L$. Park and
Plenio \cite{PP:00} show that they can implement the algorithm by use of $1$ qubit \footnote{Throughout this section, we also do not 
count the number of qubits, $\log N$ to be exact, required by the eigenvector of the
unitary.} along with the semiclassical approach \cite{GN:96}.
For such a design, the whole circuit (quantum-wise) consists of $2L + 1$ stages
of recovering $\varphi_i$, where $ 1 \leq i \leq 2L + 1$,
and calculating a controlled rotation for the next stage. After
obtaining all the $\varphi_i$, the post processing (continued fractions) recovers the order $r$.\\

In the work by Seifert\cite{Seifert:01}, he proposes an alternative to
approximate the order by using the simultaneous diophantine approximation
\cite{Lagarias:85}. The theorem is as follows. 

\begin{theorem}\cite{Seifert:01}
Let $N$ be the product of two randomly chosen primes of equal size, i.e. of the
same length in the binary representations. There exists a randomized
polynomial-time quantum algorithm that factors $N$ and uses
quantum registers of binary length $\lceil (1 + \epsilon) \log N \rceil$, where $\epsilon$ is an
arbitrarily small positive constant \footnote{$\epsilon$ determines the
dimension $d = \frac{1}{1-\frac{1}{1+\epsilon}}$ needed for the good
simultaneous diophantine approximation. It is shown
\cite{Lagarias:85} that the complexity is upper bounded from above by
$O(L^{12})$ independent of the dimension $d$.}.
\end{theorem}

In such a design, more computations are shifted from the quantum
computation part to the classical computation part, in comparison to Shor's
algorithm. This might be of importance to practical realizations of a
quantum computer. It is also clear that the simultaneous diophantine approach
only requires $(1+\epsilon)\log N$ qubits, that is the precision requirement of 
$ \frac{1}{2^{L(1+\epsilon)}}$ for the phase estimation, to guarantee the existence of a polynomial quantum algorithm for the
factorization problem. \\

Given the constraint that we only have $k$ qubits available for implementation,
we have the following scheme. 

\begin{algorithm}[H]\caption{Factorization: Choice of Approximation
Approach}\label{alg:approx_choice}
\textbf{Input:} 
			$k$ ancillary bits initialized in $|0\>$ and eigenvector $|u\>$ of unitary
			$U$ where $U|u\> = e^{2 \pi i \varphi}|u\>$. \\
    		
    		\textbf{Case I: Continued Fractions}\\
    		Choose $n = 2L + 1$. \\
    		Run algorithm \ref{alg:KQubits} to approximate $\varphi$ and the number of
    		stages is $\lceil \frac{2L+1}{k}\rceil$. \\
    		Run the continued fractions algorithm to recover the order $r$ from the
    		approximated $\varphi$.	\\
    		
    		\textbf{Case II: Diophantine Approximation} \\
    		Choose $n = L(1+\epsilon)$. \\
    		Run algorithm \ref{alg:KQubits} to approximate $\varphi$ and the number of
    		stages is $\lceil \frac{L(1+\epsilon)}{k}\rceil$. \\
    		Run the simultaneous diophantine approximation algorithm to recover the
    		order $r$ from the approximated $\varphi$.	\\
	\end{algorithm}  

Clearly this is the tradeoff between the computational complexity (even though both
are polynomial) and the available qubits. Quantumly they both have the same
number of invocation of the unitary $U$. However,  based on Eq.~\ref{eqn:invkcost}, 
the number of total quantum rotation gates invocation in
the first case is approximately $(2L + 1) \log k + (2L+1) - k$ while that of the second case is approximately $ L(1 + \epsilon) \log k + L(1+\epsilon) - k$. At the early
stage of a quantum computing implementation, $k$ is probably significantly
less than $L$. In such a scenario, the number of total rotation gate invocation
in the first case is approximately twice of that in the second case. \\

Furthermore, another important issue we need to consider is the decoherence. Despite the fact
that the complexity for case I is smaller (classically), it is more costly
quantumly. The difference in quantum resources might be amplified when the
implementation of error correction is considered as the first case has more stages and more
rotation gate invocations.

\section{Conclusion}\label{sect:theConclusion}
We expect the cost of classical computation to be fairly inexpensive in comparison to its quantum counterpart. Our approach provides a way to obtain the
eigenphase when the number of available qubits is rather limited. It
invokes $O(n \log k)$ rotation gates and this gain comes from (1) the use of recursive circuits for 
$QFT^{\dagger}$ and (2) the use of the classical bits and classical operators. \\	

Another obstacle of high-precision rotation gates (phase shift operators) is not addressed yet. For future work, 
we could combine with another approach \cite{HC:11} to approximate the eigenphase with variable number of 
qubits and arbitrary constant-precision operators.
\section{Acknowledgments}
		C.~C gratefully acknowledges the support of Lockheed Martin Corporation
		and NSF grants CCF-0726771 and CCF-0746600. We would like to thank David 
		Poulin and Pawel Wocjan for useful comments and suggestions.

	 \end{document}